\documentclass[aps,prb,twocolumn,floatfix]{revtex4-2}
\usepackage{tabularx} 
\usepackage{amsmath}  
\usepackage{amssymb}
\usepackage{graphicx} 
\usepackage{svg}
\usepackage[final]{hyperref} 
\usepackage{multirow}
\usepackage{leftindex}
\hypersetup{
	colorlinks=true,       
	linkcolor=blue,        
	citecolor=blue,        
	filecolor=magenta,     
	urlcolor=blue         
}
\usepackage{blindtext}
\usepackage{verbatim}
\usepackage{siunitx}
\usepackage[normalem]{ulem}

\begin{document}
\title{Giant strain-induced spin splitting effect in MnTe, a $g$-wave altermagnetic semiconductor}
\author{K. D. Belashchenko}
\affiliation{Department of Physics and Astronomy and Nebraska Center for Materials and Nanoscience, University of Nebraska-Lincoln, Lincoln, Nebraska 68588, USA}
\date{\today}

\begin{abstract}
Hexagonal MnTe is an altermagnetic semiconductor with $g$-wave symmetry of spin polarization in momentum space. In the nonrelativistic limit, this symmetry mandates that electric current flowing in any crystallographic direction is unpolarized. However, it is shown that elastic strain is effective in inducing the spin splitting effect in MnTe. For this analysis, a spin-orbit-coupled $\mathbf{k}\cdot\mathbf{p}$ Hamiltonian for the valence band maximum at the A point is derived and fitted to eigenvalues calculated from first principles. The spin splitting angle is calculated using the Boltzmann approach in the relaxation-time approximation. The spin splitting gauge factor exceeds 30 near the valence band maximum. Thus, with suitable substrate engineering, MnTe can be used as an efficient source and detector of spin current in spintronic devices. Proper inclusion of the Rashba-Dresselhaus spin-orbit coupling is crucial for the correct description of the transport properties of MnTe.
\end{abstract}

\maketitle

Considerable interest has recently been attracted to magnets which, considered in the nonrelativistic limit, are mandated by symmetry to have finite momentum-dependent spin splitting but vanishing magnetization \cite{Hayami2019,Yuan2020,Smejkal1,Smejkal2,Mazin-editorial}. Some promising spintronic applications of such altermagnets \cite{Smejkal1,Smejkal2} are based on the generation and detection of spin currents, which can flow both parallel and perpendicular to the charge current and lead to effects analogous to giant and tunnel magnetoresistance \cite{Shao2021}. The pure transverse spin current, called the spin splitting effect \cite{SpinSplitting-Gonzalez,SpinSplitting-Bai,SpinSplitting-Karube}, could be suitable for the switching of perpendicular magnetization in a geometry similar to spin-orbit-torque \cite{Machon-RMP} devices.

Most of the known candidates of altermagnets have collinear magnetic order, perhaps with a small relativistic spin canting. 
Among the small number of viable materials, hexagonal $\alpha$-MnTe \cite{Gonzalez2023,Osumi2024,Krempasky2024,Lee2024} is particularly promising. It has NiAs-type centrosymmetric crystal structure formed by ABAC hexagonal layer stacking, where A layers are occupied by Mn atoms. The N\'eel temperature is about 310 K \cite{Neel}, and the large Mn local moments in the alternating A layers are oriented parallel and antiparallel to a $\langle1\bar100\rangle$ in-plane direction \cite{MnTe-MCA}. Importantly for applications, MnTe is naturally hole-doped with reasonable transport properties which may be modified by doping \cite{Wasscher-thesis,Basit_RSC_2018,Basit-JAC_2019,Basit_AEM_2024}.

In terms of its nonrelativistic spin symmetry, MnTe belongs to the ${}^26/{}^2m{}^2m{}^1m$ spin Laue group, which has a $g$-wave spin-momentum profile in reciprocal space \cite{Smejkal1}. This symmetry forces the entire spin conductivity tensor to vanish. Thus, it may seem that MnTe is not suitable for spin current generation and detection in spintronic devices. The analysis presented below shows this not to be the case.  First, spin-orbit coupling effectively reduces the spin-momentum profile to $d$-wave in the $yz$ plane, where $\hat y$ is the direction of the N\'eel order parameter $\mathbf{n}$ in the standard crystallographic setting. Unfortunately, the spin splitting effect resulting from spin-orbit coupling alone turns out to be small in MnTe. However, shear strain in the $xy$ plane is effective in inducing $d$-wave spin-momentum pattern and a large strain-induced spin splitting effect. Such strain may be engineered by growing the MnTe film epitaxially on a suitable substrate, facilitating spintronic applications. In the rest of this Letter, I will describe the spin-orbit-coupled structure of the valence band maximum (VBM), fit it to first-principles data, and demonstrate the giant strain-induced spin splitting effect in MnTe.

A minimal spin-orbit-coupled single-band tight-binding model for MnTe structure was proposed in \cite{roig2024minimal}. However, the VBM at the A point \cite{Osumi2024,Faria2023} in MnTe is dominated by tellurium $p_x$ and $p_y$ orbitals \cite{Faria2023}. A realistic two-band $\mathbf{k}\cdot\mathbf{p}$ model \cite{YuCardona} was obtained in \cite{Faria2023} where a large spin-orbit splitting at the VBM was noted for $\mathbf{n}\parallel\hat z$ and described by a simple momentum-independent term. When $\mathbf{n}$ lies along the $\hat y$ easy axis, the states near the VBM are spanned by a basis of four nearly-degenerate states including a twofold orbital degree of freedom and spin \cite{Faria2023}. We will now see that spin-orbit coupling introduces several Rashba-Dresselhaus terms that are linear in momentum and important for the correct description of the transport properties of $p$-type MnTe. Note that Rashba-Dresselhaus effects at the conduction band minimum (K point) have been previously discussed \cite{RashbaK}.

We start with a basis of eight states built from the direct product of spin, orbital doublet transforming as $(p_x, p_y)$, and a sublattice-like variable represented by two nonbonding phase sequences $++--$ and $+--+$ on the Te atoms in the unit cell doubled along the $c$ axis. These nonbonding orbitals encode the correct phases at the A point. Without magnetic order, there is an eightfold degeneracy; it is lifted by spin-dependent hopping through the interleaving Mn layers, because the bonding/antibonding character of the phases is correlated with the magnetic moments of the interleaving Mn atoms. This correlation is the origin of altermagnetic spin splitting near the A point in MnTe.

Let Pauli matrices $\sigma_\alpha$, $\tau_\alpha$, and $\nu_\alpha$ act on the spin, orbital pseudospin, and sublattice pseudospin variables. First, we deduce the transformation properties of these operators and of the wavevector under the action of the nonmagnetic $P6_3/mmc$ space group combined with time-reversal operation. We form direct-product tensors up to second order in wavevector $k_\alpha$ and up to first order in each operator type, symmetrize over the space group, and read off all symmetry-allowed terms transforming as the identity representation, in the nonmagnetic state. Only terms up to first order in $k_\alpha$ are retained for spin-orbit coupling. Then the analysis is repeated to obtain additional \emph{nonrelativistic} terms allowed in the ordered altermagnetic state. To this end, we use the Shubnikov (antisymmetry \cite{heesch1930xix,Shubnikov1951}) space group \cite{Belov1955,Zamorzaev1957,Wills_2017} (also known, for collinear antiferromagnets, as the nontrivial spin group \cite{LITVIN1974538,Litvin:a14103,Smejkal1}) treating spin as a black-and-white variable to encode the magnetic order, augmented by quantum-mechanical time-reversal symmetry that has no effect on spin. This procedure gives the following nonrelativistic operators:
\begin{align}
    &\{1, k_x^2+k_y^2, k_z^2\} \{1, \sigma_n\nu_z\}\\
    \label{one}
    &(k_x\tau_x+k_y\tau_z)\sigma_n\nu_y\\
    &\left[(k_x^2-k_y^2)\tau_z+2k_xk_y\tau_x\right]\{1,\sigma_n\nu_z\}\\
    &k_z\nu_y\\
    &k_z(k_x\tau_x+k_y\tau_z)\nu_z\\
    &k_z(k_x\tau_x+k_y\tau_z)\sigma_n
\end{align}
and spin-orbit coupling:
\begin{align}
    &\tau_y\sigma_z\{1,k_z\nu_y\}\\
    &(k_x\tau_z-k_y\tau_x)\sigma_z\nu_x\\
    &k_z(\tau_x\sigma_y-\tau_z\sigma_x)\nu_x\\
    &(k_x\sigma_x+k_y\sigma_y)\tau_y\nu_y
    \label{last}
\end{align}
The $\sigma_n$ operator represents the ``black-and-white pseudoscalar,'' which in the full spin-orbit-coupled model is promoted to $\sigma_n=\boldsymbol{\sigma}\cdot\mathbf{n}$.

A combination of terms (\ref{one})-(\ref{last}) gives an eight-band $\mathbf{k}\cdot\mathbf{p}$ Hamiltonian, but four of the bands are deep below the Fermi level, and it is convenient to eliminate them, and also to rewrite the model with the spin quantization axis $z'$ parallel to the N\'eel vector. Thus, we transform to the spin-phase-locked basis by trading $\nu$ for a new pseudospin variable $\xi=2\nu\sigma$, which is acted upon by Pauli matrices $\xi_\alpha$. Further, we specialize to the case $\mathbf{n}\parallel\hat y$, L\"owdin-downfold the Hilbert space sector corresponding to $\xi=-1/2$ (which forms a quadruplet at the A point lying far below the Fermi level), and again retain spin-orbit coupling terms only up to first order in momentum. The technical details of this derivation are given in Appendix A. The final $\mathbf{k}\cdot\mathbf{p}$ model for the VBM quadruplet at the A point is
\begin{widetext}
\begin{align}
    \hat H_h = &-a_\parallel k_\parallel^2 - a_z k_z^2
+t_\Delta \left[(k_x^2-k_y^2)\tau_z+2k_xk_y\tau_x\right]
+t_z k_z \sigma_z (k_x\tau_x+k_y\tau_z)\nonumber\\
&+\lambda_1 (k_x\tau_z-k_y\tau_x)\sigma_y + \lambda_2 k_z\tau_z\sigma_x+\lambda_3 k_z\tau_y\sigma_x+\lambda_4 k_x\tau_y\sigma_y+(\Delta_0+\Delta_s)\tau_z
\label{hh}
\end{align}
\end{widetext}
where the Pauli matrices $\tau_\alpha$ and $\sigma_\alpha$ represent the orbital doublet and pseudospin degrees of freedom. Note that $\sigma_z$ is the true spin operator in the new basis, but the transverse pseudospin components $\sigma_x$, $\sigma_y$ are not. The $x$ and $y$ components of the true spin have no matrix elements within the VBM quadruplet described by Eq. (\ref{hh}), which implies that the states near the VBM have no transverse spin texture \cite{SM,Faria2023}.

The first three terms in  (\ref{hh}) describe a pair of spin-independent hole bands with in-plane effective masses $a_\parallel\pm t_\Delta$. These terms already appeared in Ref. \cite{Faria2023}. Altermagnetic spin-dependent hybridization is encoded by the $t_z$ term, which adds spin-dependent three-fold warping to the isoenergetic surfaces. The four $\lambda_i$ terms ($i=1\cdots4$) include all spin-orbit coupling up to first order in $\mathbf{k}$ that is allowed with unbroken time-reversal symmetry (i.e., in the nonmagnetic state). These terms are, however, specific to the case $\mathbf{n}\parallel\hat y$, because they are written for the basis set that is aware of the N\'eel order alignment. The tiny $\Delta_0$ term is responsible for splitting the degeneracy at the A point into two doublets separated by $2\Delta_0$ \cite{Faria2023}; this term is only allowed once the time-reversal symmetry is broken by $\mathbf{n}$ and reflects the small orthorhombic distortion of the charge density induced by the magnetic order. The $\Delta_0$ term slightly lifts the degeneracy of the tellurium $p_x$ and $p_y$ orbitals. We will return to the $\Delta_s$ term below.

I obtained the parameters of the model (\ref{hh}) by fitting to the four uppermost valence eigenvalues calculated here in VASP \cite{VASP1,VASP2,VASP3}, using the PBE$+U$ method \cite{PBE,Dudarev1998} with $U=4.8$ eV, $J=0.8$ eV \cite{Mu2019,Mazin2023}, at 803 $\mathbf{k}$ points inside a sphere of a 0.1 \AA$^{-1}$ radius around the A point. The fitting was performed by minimizing the mean-squared eigenvalue misfit using the coordinate descent method; a Gaussian weighting factor $\exp(-E^2/2\sigma_E^2)$ with a width $\sigma_E=0.15$ eV was added to emphasize the states near the VBM. The resulting parameters are as follows:
$a_\parallel=\qty{17.96}{\electronvolt\angstrom^2}$, $a_z=\qty{3.346}{\electronvolt\angstrom^2}$, $t_\Delta=\qty{10.75}{\electronvolt\angstrom^2}$, $t_z=\qty{-7.967}{\electronvolt\angstrom^2}$, $\lambda_1=\qty{0.445}{\electronvolt\angstrom}$, $\lambda_2=\qty{-0.031}{\electronvolt\angstrom}\approx0$, $\lambda_3=\qty{0.424}{\electronvolt\angstrom}$, $\lambda_4=\qty{-0.236}{\electronvolt\angstrom}$, $\Delta_0=\qty{1.28}{\milli\electronvolt}$. Some combinations of signs may be flipped without affecting the eigenvalues.

\begin{figure*}[htb]
    \centering
    \includegraphics[width=0.45\textwidth]{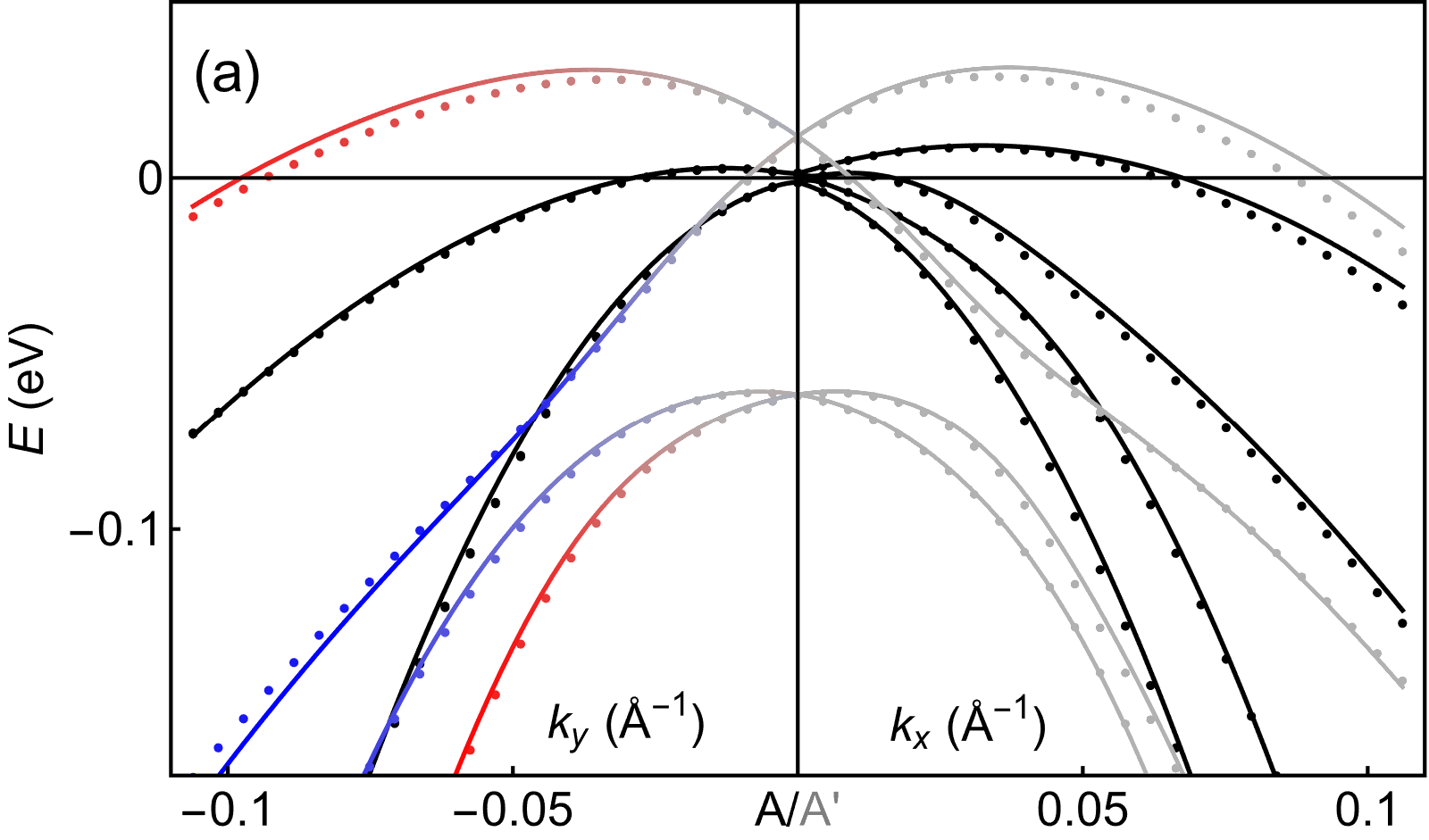}\hfil
    \includegraphics[width=0.45\textwidth]{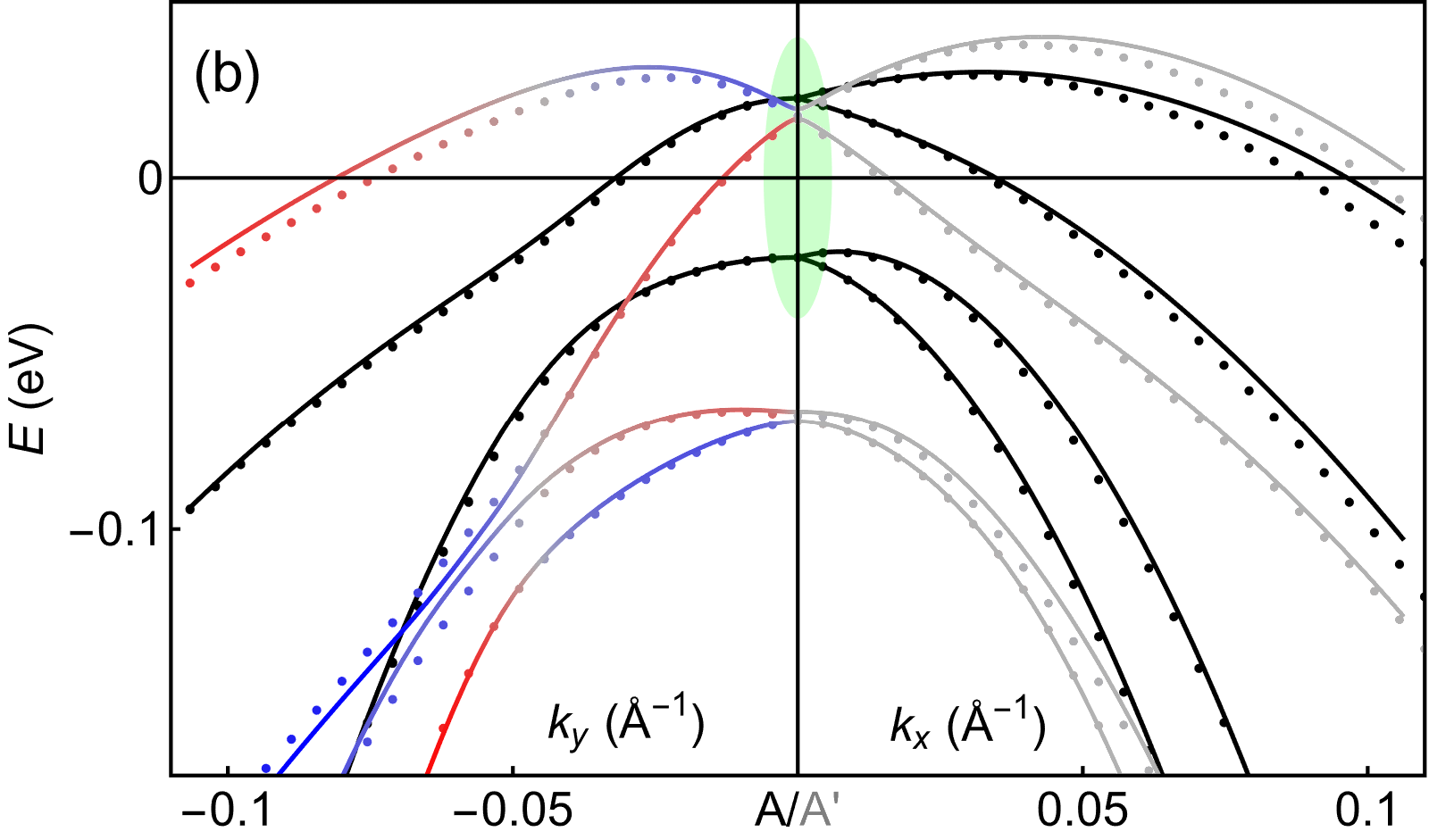}
    \caption{Relativistic band dispersion of the four VBM bands near the A point for $\mathbf{n}\parallel\hat y$. (a) Unstrained structure. (b) Strained with $\varepsilon=0.01$. Black dots and curves: unpolarized bands along the $(0,k_y,\pi/c)$ and $(k_x,0,\pi/c)$ lines. Dots and curves shaded with blue, red, and gray: bands along the $(0,k_y,\pi/c+\delta_k)$ and $(k_x,0,\pi/c+\delta_k)$ lines, where $\delta_k=0.0862$ \AA$^{-1}$. Dots: first-principles eigenvalues from VASP. Curves: eigenvalues of the fitted Hamiltonian (\ref{hh}). Red (blue) shading: positive (negative) spin polarization of the eigenstates at $k_z=\pi/c+\delta_k$; unpolarized states appear light gray. The green ellipse highlights the strain-induced splitting $\Delta_s$ at the A point.
    }
    \label{fig:fit}
\end{figure*}

The band dispersions near the VBM and the quality of the fit are illustrated by Fig. \ref{fig:fit}(a). The spin-orbit coupling lifts the degeneracy almost everywhere on the (otherwise nodal) $k_z=\pi/c$ plane. Hovever, the bands remain doubly degenerate along the nodal line $\mathbf{k}=(0,k_y,\pi/c)$. There are two additional nodal lines (not shown in Fig. \ref{fig:fit}) in the $k_z=\pi/c$ plane which are at $\pm\pi/6$ angles to the $k_x$ axis if $\Delta_0$ is set to zero but are otherwise slightly tilted. Note, however, that, even at $\Delta_0=0$, the $C_{3z}$ symmetry axis in the eigenvalue spectrum is broken by the $\lambda_4$ term at any $k_z$ and also by the $\lambda_2$ term at finite $k_z$. At finite $k_z$, the degeneracy is lifted by the altermagnetic $t_z$ term at all finite $\mathbf{k}_\parallel$ [see Fig. \ref{fig:fit}(a)].

The spin polarization vanishes on the $k_z=\pi/c$ and $k_y=0$ planes, but it is finite on the $(0,k_y,\pi/c+\delta_k)$ line shown in Fig. \ref{fig:fit}(a). The spin polarization of the first-principles eigenstates is obtained by taking the sum of spin projection on the atomic spheres and normalizing it to its maximum value over the whole Brillouin zone for the four relevant bands. This spin polarization, shown in red and blue colors in Fig. \ref{fig:fit}(a), agrees well with that of the fitted Hamiltonian (\ref{hh}).

The parameter $\Delta_s$ may be obtained from the strain-induced crystal field splitting at the A point.
Figure \ref{fig:fit}(b) shows the band dispersions calculated from first principles for the volume-conserving biaxial strain $\epsilon_{xx}=-\varepsilon_{yy}=\varepsilon/2$ with $\varepsilon=0.01$; the crystal field splitting of the states at $k_z=\pi/c$ is highlighted by the green ellipse. The band dispersions obtained with the term $\Delta_s=21.3$ meV added to the fitted unstrained Hamiltonian in Eq. (\ref{hh}) are also shown in Fig. \ref{fig:fit}(b). Although strain also makes other terms in Eq. (\ref{hh}) anisotropic, the good fit in Fig. \ref{fig:fit}(b) allows us to ignore this anisotropy and describe the effect of the biaxial strain, to linear order, by a single term $\Delta_s=2.13\varepsilon$ eV. Note the agreement in the spin polarization of the eigenstates along the $(0,k_y,\pi/c+\delta_k)$ line in Fig. \ref{fig:fit}(b), including the sign inversions in all four bands, despite the fact that the fitting procedure ignored the spin polarization.

\begin{figure*}[htb]
    \centering
\includegraphics[width=0.95\textwidth]{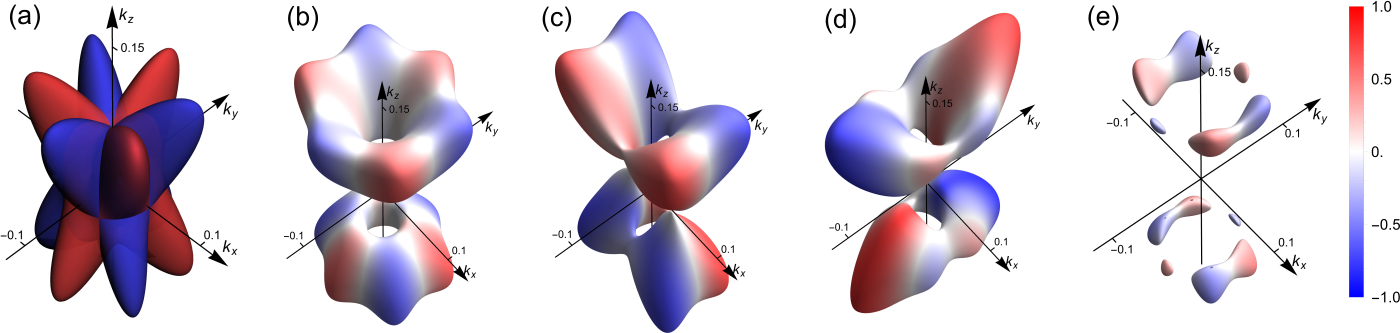}
    \caption{(a-d) Isosurfaces in MnTe at hole concentration $n=10^{-3}$ per unit cell. (a) Without spin-orbit coupling. Red (blue) color: electrons with spin up (down). (b) With spin-orbit coupling and $\mathbf{n}\parallel\hat y$. Color represents the spin polarization along $\mathbf{n}$ as shown by the color bar, where $\pm1$ stands for $\langle\mathbf{sn}\rangle=\pm1/2$.
    (c) Same as (b) but with added volume-conserving shear strain $\varepsilon_{xx}=-\varepsilon_{yy}=\varepsilon/2$ with $\varepsilon=0.01$. (d) Same as (c) but with $\varepsilon=-0.01$. (e) Isosurface at 1 meV below VBM showing its valley structure.}
    \label{fig:iso}
\end{figure*}

Let us now examine the isoenergetic surfaces and momentum-dependent spin polarization near VBM. We fix the hole concentration at $10^{-3}$ holes per unit cell (i.e., $10^{19}$ cm$^{-3}$, easily achievable by doping \cite{Wasscher-thesis,Basit_RSC_2018,Basit-JAC_2019,Basit_AEM_2024}).
Figure \ref{fig:iso}(a) shows the isosurface of the fitted Hamiltonian (\ref{hh}) with the spin-orbit coupling switched off. The spin-up and spin-down electrons are decoupled in this case, and the corresponding isoenergetic surfaces are shown in red and blue color, respectively. These surfaces have the characteristic $g$-wave altermagnetic pattern with one nodal plane at $k_z=\pi/c$ and three vertical nodal planes.

Figure \ref{fig:iso}(b) shows an isosurface of the uppermost hole band with spin-orbit coupling fully included. Although spin-orbit coupling reduces the space group to orthorhombic $Cm'c'm$, the removal of the $C_{3z}$ axis is not visible on this isosurface; it still appears to have the altermagnetic $g$-wave pattern, albeit with a significant mixing of the spin-up and spin-down character and a strongly suppressed spin polarization on large swaths of the isosurface. The spin-orbit coupling-induced loss of the $g$-wave spin-momentum pattern is easier to see in Fig. \ref{fig:iso}(e) close to the top of the VBM, where two warped toroids have shrunk and split into a set of four larger and four smaller valleys.

Symmetry analysis of the piezo-spin-galvanic tensor (Appendix B) reveals that the spin conductivity tensor in MnTe acquires a finite $\sigma^s_{yz}$ component under $\varepsilon_{xx}$ or $\varepsilon_{yy}$ distortion, already in the nonrelativistic case.
The effect of volume-conserving shear strain of this symmetry ($\varepsilon_{xx}=-\varepsilon_{yy}=\varepsilon/2$) on the structure of the isosurfaces near the VBM is illustrated by Figs. \ref{fig:iso}(c) and \ref{fig:iso}(d), where $\varepsilon=\pm0.01$. As explained above, the strain affects the electronic structure through the $\Delta_s\tau_z$ term in the Hamiltonian (\ref{hh}).
As seen in Figs. \ref{fig:iso}(c) and \ref{fig:iso}(d), a relatively small 1\% shear strain in the $xy$ plane results in a strong distortion of the hole isosurfaces and an obvious loss of the $g$-wave altermagnetic symmetry. The spin polarization at these isosurfaces has a clearly pronounced $d$-wave symmetry in the $yz$ plane.

\begin{figure}[hbt]
    \centering
    \includegraphics[width=0.45\textwidth]{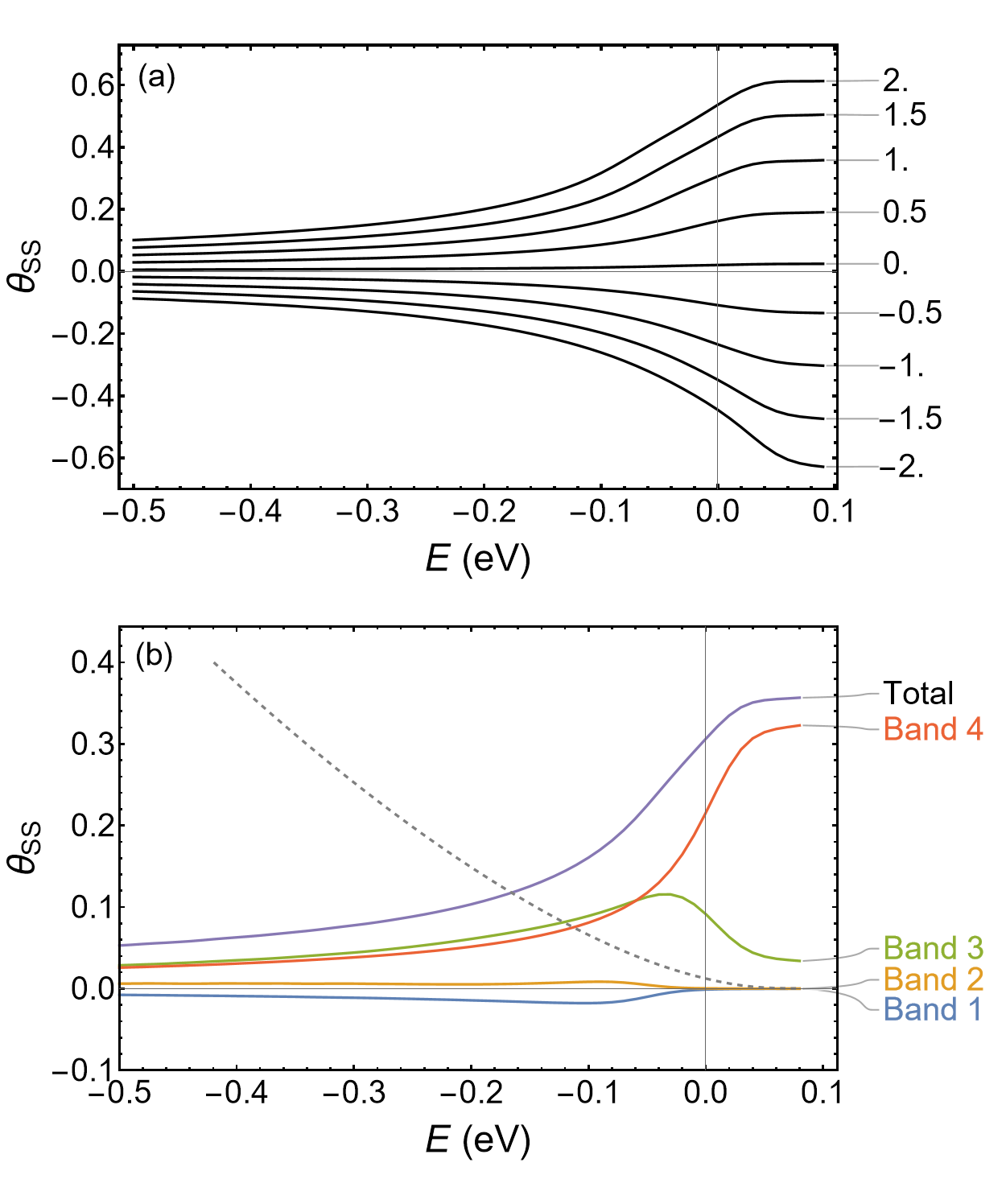}
    \caption{(a) Spin splitting angle $\theta_\mathrm{SS}$ as a function of the chemical potential near the top of the valence band in MnTe, at 150 K, for strain magnitudes $\varepsilon$ indicated on the right, in percent.
    (b) Contributions from the four hole bands, in the order of increasing energy, to $\theta_\mathrm{SS}$ at $\varepsilon=0.01$. Dashed line: $\sigma_{zz}(E)$ (arb. units).}
    \label{fig:SHE}
\end{figure}

Figure \ref{fig:SHE}(a) shows the spin splitting angle $\theta_\mathrm{SS}=\sigma^s_{xz}/\sigma_{zz}$ (defined similar to the spin Hall angle) in MnTe as a function of the chemical potential. The charge and spin conductivities are calculated using the Boltzmann approximation \cite{ziman1960book} with the transport relaxation time $\tau$ assumed to be the same for all states:
\begin{align}
    \sigma_{\alpha\beta}&=\tau\sum_n\int v_{n\alpha} v_{n\beta}\frac{\partial f(E_n)}{\partial \mu}\frac{d^3k}{(2\pi)^3}
    \label{cond}
    \\
    \sigma^s_{\alpha\beta}&=\tau\sum_n\int v_{n\alpha} v_{n\beta}s_{ny}\frac{\partial f(E_n)}{\partial \mu}\frac{d^3k}{(2\pi)^3},
    \label{scond}
\end{align}
where $\mathbf{v}_n(\mathbf{k})=\langle n|\partial\hat H_h/\partial \mathbf{k}|n\rangle$ and $s_{ny}(\mathbf{k})=\langle n|\sigma_z|n\rangle$ are the group velocity and spin expectation value along the $\mathbf{n}\parallel\hat y$ axis for the eigenstate $|n\rangle$, $E_n(\mathbf{k})$ is its energy eigenvalue, and $f(E)$ the Fermi-Dirac distribution function. Integration in Eqs. (\ref{cond})-(\ref{scond}) was performed numerically using a dense mesh around the A point at a Fermi temperature $T=150$ K. (The results for $T=300$ K are similar, with $\theta_\mathrm{SS}$ reduced by less than 10\%.)

Figure \ref{fig:SHE}(a) indicates that a relatively small shear strain induces a large spin splitting effect, which is consistent with our expectations based on Figs. \ref{fig:iso}(c-d). The dimensionless spin splitting \emph{gauge factor} $\theta_\mathrm{SS}/\varepsilon$ at $\mu=0$ is more than 30 reflecting a strong sensitivity of the Fermi surface and spin polarization anisotropy to shear strain. 
Figure \ref{fig:SHE}(b) resolves the contributions of the four hole bands to $\sigma^s_{xz}$ at $\varepsilon=0.01$. We see that the spin splitting effect is dominated by the spin conductivity of the top two hole bands. Note that the isosurfaces seen in Fig. \ref{fig:iso} correspond to the uppermost Band 4 in Fig. \ref{fig:SHE}.

If spin-orbit coupling is turned off, the dependence of the spin splitting angle on the chemical potential remains qualitatively similar, but the gauge factor at $E\approx0$ exceeds 100. This drastic difference highlights the importance of including spin-orbit coupling for the correct description of the transport properties in $p$-type MnTe.

Figure \ref{fig:SHE}(a) also shows that $\theta_\mathrm{SS}$ is small but finite (about 0.02 at $E=0$) at zero strain due to the symmetry reduction introduced by spin-orbit coupling. Note that this symmetry reduction occurs even when $\Delta_0=\Delta_s=0$ and the charge density retains the full nonmagnetic space group symmetry. An analysis based on the nonrelativistic spin group does not account for this symmetry reduction and predicts a strictly zero $\theta_\mathrm{SS}$.

Using the general form of spin-orbit coupling for $\mathbf{n}\perp\hat z$ (Appendix A), I have found that the isosurfaces shown in Fig. \ref{fig:iso}(b-d) and the strain-induced spin splitting effect are almost independent of the in-plane orientation of the order parameter $\mathbf{n}$.

In order to exploit the large strain-induced spin splitting effect in MnTe for spintronics applications, the MnTe crystal needs to be grown on a substrate imposing a shear strain in the crystallographic $ab$ plane. The likely geometry would have the charge current flowing along the hexagonal axis lying in the film plane, and the spin current flowing perpendicular to the film plane (see Appendix C for further discussion). The spin current is polarized along the N\'eel order parameter, which could have a normal component suitable for the switching of perpendicular magnetization.

\begin{acknowledgments}

I thank Vladimir Antropov, Igor Mazin, and Alexey Kovalev for useful comments about the manuscript. This work was supported by the U.S. Department of Energy (DOE) Established Program to Stimulate Competitive Research (EPSCoR) grant No. DE-SC0024284. Some calculations were performed utilizing the Holland Computing Center of the University of Nebraska, which receives support from the Nebraska Research Initiative.

\end{acknowledgments}

\appendix

\section{$\mathbf{k}\cdot\mathbf{p}$ Hamiltonian for M\lowercase{n}T\lowercase{e}}
\label{kp_MnTe}

As explained in the main text, we first obtain an 8-band $\mathbf{k}\cdot\mathbf{p}$ model for the vicinity of the A point, which includes the following terms transforming as the identity representation of the MnTe space group:
\begin{align}
    &\{1, k_x^2+k_y^2, k_z^2\} \{1, \sigma_n\nu_z\}\\
    &(k_x\tau_x+k_y\tau_z)\sigma_n\nu_y\\
    &\left[(k_x^2-k_y^2)\tau_z+2k_xk_y\tau_x\right]\{1,\sigma_n\nu_z\}\\
    &k_z\nu_y\\
    &k_z(k_x\tau_x+k_y\tau_z)\nu_z\\
    &k_z(k_x\tau_x+k_y\tau_z)\sigma_n\\
    &\tau_y\sigma_z\{1,k_z\nu_y\}\quad\textrm{(SOC)}\\
    &(k_x\tau_z-k_y\tau_x)\sigma_z\nu_x \quad\textrm{(SOC)}\\
    &k_z(\tau_x\sigma_y-\tau_z\sigma_x)\nu_x \quad\textrm{(SOC)}\\
    &(k_x\sigma_x+k_y\sigma_y)\tau_y\nu_y\quad\textrm{(SOC)}
\end{align}
where $\sigma_n=\boldsymbol{\sigma}\cdot\mathbf{n}$ and spin-orbit coupling terms are notated as ``SOC.'' Curly brackets denote sets of terms, any of which can appear as a factor. The terms that do not contain any $\sigma$ matrices are allowed by the nonmagnetic space group, regardless of whether spin is treated relativistically or as a black-and-white variable. The SOC terms are allowed by the nonmagnetic space group with the spin treated relativistically. The terms containing $\sigma_n$ are allowed by the Shubnikov antisymmetry space group \cite{Shubnikov1951,Belov1955,Zamorzaev1957} in the altermagnetically ordered state treating spin as a black-and-white variable.

Next, we choose a reference frame with the spin quantization axis $z'$ parallel to the N\'eel order parameter $\mathbf{n}$. We switch to the new basis labeled by $\xi$ and $\sigma$ where $\xi=2\nu\sigma$, which is acted upon by Pauli matrices $\xi_\alpha$. In other words, the new basis states are $(\xi,\sigma=1/2)\equiv(\nu=\xi,\sigma=1/2)$, 
$(\xi,\sigma=-1/2)\equiv(\nu=-\xi,\sigma=-1/2)$. In this new basis, the spin and sublattice operators are represented as follows:
\begin{align}
    &\{\sigma_{x'},\sigma_{y'},\sigma_{z'}\}\to\{\xi_x\tilde\sigma_{x'},\xi_x\tilde\sigma_{y'},\sigma_{z'}\}\label{spin-op}\\
    &\{\nu_x,\nu_y,\nu_z\}\to\{\xi_x,\xi_y\sigma_{z'},\xi_z\sigma_{z'}\}
\end{align}
The primes remind us that the spin matrices are defined in the primed reference frame where $\hat z'\parallel\mathbf{n}$. We drop these primes in the following.

I emphasize that while the $\sigma_z$ operator still represents true spin in the new basis, the operators $\tilde\sigma_x$, $\tilde\sigma_y$ do not. The transverse components of the true spin operator [left-hand side of Eq. (\ref{spin-op})] include the $\xi_x$ matrix in the new basis [right-hand side of Eq. (\ref{spin-op})]. The tilde signs are omitted in the following.

The transformation to the new basis results in the following terms:
\begin{align}
    &\{1, k_x^2+k_y^2, k_z^2\} \{1, \xi_z\}\\
    &(k_x\tau_x+k_y\tau_z)\xi_y\\
    &\left[(k_x^2-k_y^2)\tau_z+2k_xk_y\tau_x\right]\{1,\xi_z\}\\
    &k_z\xi_y\sigma_z\\
    &k_z(k_x\tau_x+k_y\tau_z)\xi_z\sigma_{z}\\
    &k_z(k_x\tau_x+k_y\tau_z)\sigma_{z}\\
    &\tau_y\sigma'_z\{1,k_z\xi_y\sigma_z\} \quad\textrm{(SOC)}\\
    &(k_x\tau_z-k_y\tau_x)\xi_x\sigma'_z \quad\textrm{(SOC)}\\
    &k_z(\tau_x\sigma'_y-\tau_z\sigma'_x)\xi_x \quad\textrm{(SOC)}\\
    &(k_x\sigma'_x+k_y\sigma'_y)\tau_y\xi_y\sigma_z\quad\textrm{(SOC)}
\end{align}
where
\begin{align}
\sigma'_\alpha = R_{\alpha x'}\xi_x\sigma_{x}+R_{\alpha y'}\xi_x\sigma_{y}+R_{\alpha z'}\sigma_{z}
\end{align}
and the matrix $R_{\alpha\beta}$ contains the components of the ``old'' unit vectors along the crystallographic axes $\alpha$ in the new reference frame where $\mathbf{n}\parallel\hat z'$. Simply put, $\sigma'_\alpha$ is the representation of the original $\sigma_\alpha$ spin matrix in the new basis set corresponding to a given direction of $\mathbf{n}$. Note the presence of the $\xi_x$ matrix in the definition of $\sigma'_\alpha$. Also note the difference between $\sigma_\alpha$ and $\sigma'_\alpha$: the former are already defined in the reference frame where $\mathbf{n}\parallel z'$, but the latter are defined in the original, crystallographic reference frame.

To clarify the meaning of the transformation, consider two cases: (i) $\mathbf{n}=\hat z$, and (ii) $\mathbf{n}=\hat y$. In case (i), $R_{\alpha\beta}=\hat 1$ is a unit matrix, and the primed spin operators in the SOC terms are equal to their usual unprimed counterparts. In case (ii), we may choose the $R_{\alpha\beta}$ matrix to represent a rotation about the $\hat x$ axis by an angle $\pi/2$, whereby $\sigma'_{x}=\xi_x\sigma_{x}$, $\sigma'_{y}=\sigma_{z}$, and $\sigma'_{z}=-\xi_x\sigma_{y}$.

We now assume that the $\xi_z$ term in the Hamiltonian comes with a positive coefficient and downfold the Hilbert space sector corresponding to $\xi=-1/2$, which forms a quadruplet at the A point lying deep below the Fermi level. This downfolding will result in an effective Hamiltonian for the quadruplet of bands at the valence band maximum (VBM).

As emphasized above, the transverse components of the true spin operator (\ref{spin-op}) include the off-diagonal $\xi_x$ matrix. This means that any linear combination of the four VBM basis states with $\xi=1/2$ has zero expectation values of the transverse true spin components. In other words, apart from the small admixture of the states outside of the VBM quadruplet, the states near the top of the valence band exhibit no transverse spin texture. This feature was noted in Ref. \cite{Faria2023}.

Let us specialize to the case $\mathbf{n}\perp\hat z$ where the order parameter lies in the $xy$ plane. The SOC operators become
\begin{align}
    &\tau_y\xi_x\sigma_y\{1,k_z\xi_y\sigma_z\}\\
    &(k_x\tau_z-k_y\tau_x)\sigma_y \\
    &k_z[(\tau_x n_y-\tau_z n_x)\sigma_z\xi_x-(\tau_x n_x+\tau_z n_y)\sigma_x]\\
    &[(k_x n_y - k_y n_x) \xi_z\sigma_y+(k_x n_x + k_y n_y)\xi_y]\tau_y
\end{align}
The first, $k$-independent term has no matrix elements within the VBM quadruplet thanks for the presence of the $\xi_x$ matrix. As a result, for $\mathbf{n}\perp\hat z$ the fourfold degeneracy at the A point is not lifted by this SOC term.

To first order in $k$, downfolding does not add any new terms to the VBM sector, and we are left only with terms that are diagonal in $\xi$:
\begin{align}
    &k_z\tau_y\sigma_x\\
    &(k_x\tau_z-k_y\tau_x)\sigma_y \\
    &k_z(\tau_x n_x + \tau_z n_y)\sigma_x\\
    &(k_x n_y - k_y n_x) \sigma_y\tau_y
\end{align}

Choosing $\mathbf{n} \parallel \hat y$, which is an easy axis, this results in the Hamiltonian (11) of the main text sans the $\tau_z$ term. That $\tau_z$ term is allowed by the relativistic magnetic space group $Cm'c'm$ and reflects the effect of the N\'eel order parameter $\mathbf{n}\parallel\hat y$ on the charge density.

\section{Piezo-spin galvanic tensor}

To describe the effect of elastic strain on the spin conductivity, it is convenient to introduce the \emph{piezo-spin-galvanic} tensor $P_{ijkl}$ as follows:
\begin{equation}
    \sigma^s_{ij}=\sigma^{s0}_{ij}+P_{ijkl}\varepsilon_{kl}
\end{equation}
where $\sigma^s_{ij}=\sigma^\uparrow_{ij}-\sigma^\downarrow_{ij}$ is the spin conductivity tensor, and $\varepsilon_{kl}$ is the elastic strain tensor. 
Here we assume that the effects of spin-orbit coupling are negligible, which allows us to use the Shubnikov space group treating the spin as a black-and-white variable \cite{Shubnikov1951,Belov1955,Zamorzaev1957}. This also implies that the spin conductivity tensor $\sigma^s_{ij}$ is symmetric. The tensor $P_{ijkl}$ is, therefore, symmetric with respect to its first and second pairs of indices: $P_{ijkl}=P_{jikl}=P_{ijlk}$.

Because the tensor $P_{ijkl}$ is invariant under inversion, its structure depends only on the spin Laue group and can be established using Neumann's principle. We will limit our consideration to materials where $\sigma^{s0}_{ij}=0$, i.e., where symmetry forbids generation of spin currents by electric field in any geometry in the absence of strain.

\begin{table}[h!]
\caption{Nonzero elements (up to generic permutations) of the piezo-spin-galvanic tensor for spin Laue groups where the spin conductivity tensor vanishes at zero strain. For the ${}^26/{}^2m{}^2m{}^1m$ group, it is assumed that one of the ${}^2m$ planes is normal to the $y$ axis.}
\begin{tabular}{|l|l|}
\hline
Spin Laue group & Nonzero elements of $P_{ijkl}$\\
\hline
    ${}^14/{}^1m{}^2m{}^2m$   & $P_{xyxx}=-P_{xyyy}=a$\\
                            & $P_{xxxy}=-P_{yyxy}=b$\\
                            & $P_{xzyz}=-P_{yzxz}=c$\\
    ${}^16/{}^1m{}^2m{}^2m$   & Same as above with $a=-b$\\
    ${}^26/{}^2m$             & $P_{xzxx}=-P_{xzyy}=-P_{yzxy}=a$      \\
                            & $P_{yzxx}=-P_{yzyy}=P_{xzxy}=b$     \\
                            & $P_{xxxz}=-P_{yyxz}=-P_{xyyz}=c$ \\
                            & $P_{xyxz}=P_{xxyz}=-P_{yyyz}=d$  \\
    ${}^26/{}^2m{}^2m{}^1m$   & Same as above with $a=c=0$   \\
    ${}^1{m}{}^1{\bar3}{}^2{m}$   & $P_{xxyy}=-P_{yyxx}$ and cyclic permutations\\
    \hline
\end{tabular}

\label{tab:nonzero}
\end{table}

Table \ref{tab:nonzero} lists all nonzero elements of $P_{ijkl}$ for such spin Laue groups. The case of MnTe, which is the focus of this Letter, corresponds to spin Laue group ${}^26/{}^2m{}^2m{}^1m$. The strain-induced spin splitting effect considered in the main text corresponds to the tensor element $P_{yzxx}=b$. We focused on this element, rather than the other nonzero parameter $d$, because it has a more natural implementation using substrate engineering. The next section provides an example.

\section{Choice of a substrate}

With the hexagonal axis lying in the plane of the film, there are two low-index options for the surface normal: $\langle 10\bar10\rangle$ (Y-cut) and $\langle 2\bar1\bar10\rangle$ (X-cut), with lattice constants of $a=4.13$ \AA\ and $\sqrt{3}a=7.160$ \AA, respectively. The lattice constant along the hexagonal axis is $c=6.65$ \AA, which includes four atomic monolayers. The ratio $\sqrt{3}a/(3c/4)\approx1.436$ is 1.5\% larger than $\sqrt{2}$. Thus, a cubic $\langle110\rangle$ substrate with a lattice constant of about $3c/4\approx 5.0$ \AA\ could promote the growth of a MnTe film with a $\langle 2\bar1\bar10\rangle$ surface and a shear strain on the order of a percent. Yttria-stabilized zirconia with a lattice constant of 5.125 \AA\ could be a possible candidate.

\end{document}